\documentclass[sigconf]{acmart}

\AtBeginDocument{%
  }

\copyrightyear{2025}
\acmYear{2025}
\setcopyright{acmlicensed}\acmConference[SIGIR '25]{Proceedings of the 48th International ACM SIGIR Conference on Research and Development in Information Retrieval}{July 13--18, 2025}{Padua, Italy}
\acmBooktitle{Proceedings of the 48th International ACM SIGIR Conference on Research and Development in Information Retrieval (SIGIR '25), July 13--18, 2025, Padua, Italy}
\acmDOI{10.1145/3726302.3730061}
\acmISBN{979-8-4007-1592-1/2025/07}

\usepackage{multirow}
\usepackage{threeparttable}
\usepackage{makecell}
\usepackage{tcolorbox}
\usepackage{booktabs}
\usepackage{array}
\usepackage{colortbl}
\usepackage{xcolor}
\usepackage{subcaption}
\usepackage{enumitem}
\usepackage{bm}
\usepackage{float}
\usepackage{balance}
\usepackage{amsmath}
\usepackage{pifont}

\definecolor{color_effe}{HTML}{4e79a7}
\definecolor{color_effi}{HTML}{f28e2b}

\definecolor{pastel_blue}{RGB}{161,200,243}
\definecolor{pastel_orange}{RGB}{255,180,130}
\definecolor{pastel_green}{RGB}{141,230,161}

\begin{document}

\title{Precise Zero-Shot Pointwise Ranking with LLMs through Post-Aggregated Global Context Information}

\author{Kehan Long}
\orcid{0009-0007-0827-6211}
\affiliation{%
  \institution{National University of Defense Technology}
  \city{Changsha}
  \country{China}
}
\email{longkehan15@nudt.edu.cn}

\author{Shasha Li}
\authornote{Corresponding authors.}
\orcid{0000-0002-6508-5119}
\affiliation{%
  \institution{National University of Defense Technology}
  \city{Changsha}
  \country{China}
}
\email{shashali@nudt.edu.cn}

\author{Chen Xu}
\orcid{0009-0009-1715-435X}
\affiliation{%
  \institution{National University of Defense Technology}
  \city{Changsha}
  \country{China}
}
\email{xuchen@nudt.edu.cn}

\author{Jintao Tang}
\authornotemark[1]
\orcid{0000-0002-8802-3906}
\affiliation{%
  \institution{National University of Defense Technology}
  \city{Changsha}
  \country{China}
}
\email{tangjintao@nudt.edu.cn}

\author{Ting Wang}
\authornotemark[1]
\orcid{0000-0002-7780-2330}
\affiliation{%
  \institution{National University of Defense Technology}
  \city{Changsha}
  \country{China}
}
\email{tingwang@nudt.edu.cn}

\renewcommand{\shortauthors}{Kehan Long, Shasha Li, Chen Xu, Jintao Tang, and Ting Wang}

\begin{abstract}
Recent advancements have successfully harnessed the power of Large Language Models (LLMs) for zero-shot document ranking, exploring a variety of prompting strategies. 
Comparative approaches like pairwise and listwise achieve high effectiveness but are computationally intensive and thus less practical for larger-scale applications.
Scoring-based pointwise approaches exhibit superior efficiency by independently and simultaneously generating the relevance scores for each candidate document. 
However, this independence ignores critical comparative insights between documents, resulting in inconsistent scoring and suboptimal performance.
In this paper, we aim to improve the effectiveness of pointwise methods while preserving their efficiency through two key innovations:
(1) We propose a novel \textbf{G}lobal-\textbf{C}onsistent \textbf{C}omparative \textbf{P}ointwise Ranking (\textbf{GCCP}) strategy that incorporates global reference comparisons between each candidate and an \textit{anchor} document to generate contrastive relevance scores. 
We strategically design the \textit{anchor} document as a query-focused summary of pseudo-relevant candidates, which serves as an effective reference point by capturing the global context for document comparison.
(2) These contrastive relevance scores can be efficiently \textbf{P}ost-\textbf{A}ggregated with existing pointwise methods, seamlessly integrating essential \textbf{G}lobal \textbf{C}ontext information in a training-free manner (\textbf{PAGC}).
Extensive experiments on the TREC DL and BEIR benchmark demonstrate that our approach significantly outperforms previous pointwise methods while maintaining comparable efficiency.
Our method also achieves competitive performance against comparative methods that require substantially more computational resources.
More analyses further validate the efficacy of our \textit{anchor} construction strategy.
\end{abstract}




\maketitle

\section{Introduction}
\label{sec:intro}

\begin{figure*}[t]
    \centering
    \includegraphics[width=\textwidth]{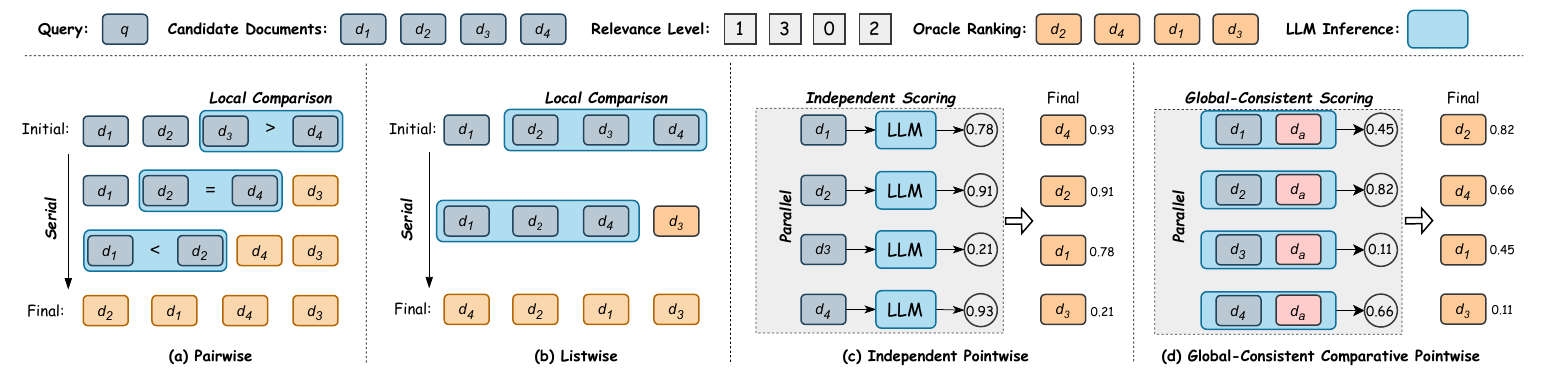}
    \vspace*{-1.8\baselineskip}
    \caption{Illustration of different prompting strategies. Comparative approaches include (a) Pairwise and (b) Listwise. Pointwise Scoring approaches include (c) Independent Pointwise and (d) our proposed Global-Consistent Comparative Pointwise. }
    \vspace*{-0.8\baselineskip}
    \label{fig:different_wise}
\end{figure*}

With the emergence of Large Language Models (LLMs), recent research has made promising advances in leveraging LLMs for IR tasks, particularly within a zero-shot paradigm~\cite{WangSKZ23, s-preference, KojimaGRMI22}.
Based on the mechanism how LLMs are used for the zero-shot document ranking task, existing methods can generally be categorized into two main types:  Comparative methods \cite{pairwise, prp-graph, lrl, rankgpt, setwise, rankVicuna} and Pointwise Scoring methods \cite{qg, qg-risk, qg-zhuang, rg-yn, rg-s}.

Comparative methods instruct LLMs to determine the relative order among documents using strategies such as pairwise and listwise prompting.
Pairwise methods (Figure~\ref{fig:different_wise}a) directly compare two documents at a time, employing voting systems like Copeland’s Method \cite{copeland} or sorting algorithms like bubble sort to establish the ranking across all candidate documents \cite{pairwise}. 
Listwise methods (Figure~\ref{fig:different_wise}b) involve a series of local comparisons within a small window, applying a sliding window strategy to construct the final global ranking \cite{rankgpt, rankVicuna}. 
Although comparative methods effectively leverage LLMs to enhance rankings, they suffer from significant challenges such as sensitivity to input order, high computational costs, and inefficiency.

In contrast, pointwise scoring methods are valued for their simplicity and high efficiency in LLM applications \cite{consolidating, s-preference, rel-judgment}. 
Current scoring methods (Figure~\ref{fig:different_wise}c) independently evaluate each document, using the probability of generating specific labels \cite{rg-yn, rg-s} or according queries \cite{qg, qg-zhuang} as relevance scores. 
However, while pointwise scoring methods offer notable efficiency, their effectiveness is hampered by two key limitations. 
Firstly, the independent scoring of each document fails to incorporate important comparative insights, often leading to inconsistencies in relevance assessments.
Secondly, there is a significant gap between the general training of LLMs and the specific calibration needed for precise ranking tasks, leaving relevance scores unrefined. 

While recent approaches \cite{setwise, tourrank, top-down, first} have explored different trade-offs between effectiveness and efficiency, no method has yet matched the efficiency of pointwise methods with the effectiveness of comparative approaches.
In this paper, we aim to achieve such performance by enhancing the effectiveness of pointwise methods from two perspectives:
Firstly, we propose a novel \textbf{G}lobal-\textbf{C}onsistent \textbf{C}omparative \textbf{P}ointwise Ranking (\textbf{GCCP}) strategy that fundamentally addresses the independence limitation of pointwise methods. 
As shown in Figure~\ref{fig:different_wise}d, GCCP introduces an \textit{anchor} document, constructed through an unsupervised spectral-based multi-document summarization approach, which serves as a global reference point for efficient yet informative comparisons. 
This \textit{anchor} document is designed to be both query-focused and globally representative, enabling more consistent relevance assessments across the document set.
Secondly, we introduce a \textbf{P}ost-\textbf{A}ggregation with \textbf{G}lobal \textbf{C}ontext (\textbf{PAGC}) framework that effectively combines contrastive relevance scores from GCCP with existing pointwise scores. 
Benefiting from the inherent efficiency of pointwise methods that assign scores to individual documents, various pointwise rankings can be efficiently combined, even through straightforward linear aggregation techniques.
Unlike the homogeneous aggregation of previous independent pointwise rankings, incorporating the global context from GCCP leads to substantial improvements in relevance assessment.
This training-free approach bridges the gap between LLM capabilities and ranking requirements while maintaining the efficiency advantages of pointwise methods.

We evaluate our GCCP and PAGC approaches along with other existing approaches on the TREC DL and BEIR benchmarks.
We conduct experiments with Flan series LLMs~\cite{flan-models} in line with a recent systematic evaluation of zero-shot LLM rankers~\cite{setwise}.
Empirical results demonstrate that our approach significantly outperforms existing pointwise methods while maintaining comparable efficiency, and achieves competitive performance against comparative methods that require substantially more computational resources. 
Moreover, both theoretical analyses and empirical results of latency and cost show the superiority of our approach in the trade-off between effectiveness and efficiency.
Further analyses validate the effectiveness of our \textit{anchor} construction strategy in enabling informative global-consistent comparisons, surpassing various intuitive strategies like synthetic documents.
Our code is available is at \url{https://github.com/ChainsawM/GCCP}.

The main contributions of this work are summarized below:
\vspace{-0.3em}
\begin{itemize}[leftmargin=*]
\item We propose a novel GCCP ranking strategy that introduces an \textit{anchor} document as a global reference point, enabling efficient yet informative comparisons while maintaining the efficiency of pointwise methods.

\item We develop a PAGC framework that effectively combines contrastive relevance scores with existing pointwise scores in a training-free manner, enhancing ranking performance while preserving efficiency.

\item Extensive experiments on TREC DL and BEIR benchmarks demonstrate that our approach significantly outperforms existing pointwise methods while maintaining comparable efficiency, and achieves competitive performance against more computationally intensive comparative methods.
\end{itemize}

\section{Related Work}
\label{sec:relatedwork}

Existing LLM-based document rankers are usually categorized into three approaches based on the number of documents simultaneously processed with a query: \textit{Pointwise} \cite{qg, qg-risk, qg-zhuang, rg-yn, rg-s}, \textit{Pairwise} \cite{pairwise, prp-graph}, and \textit{Listwise} \cite{lrl, rankgpt, setwise, rankVicuna}. 
From a novel perspective, we propose to classify current methods based on their ranking mechanisms into two general categories: \textit{Pointwise Scoring Approaches}, which employ pointwise inference where LLMs generate individual relevance scores for ranking, and \textit{Comparative Approaches}, which encompass both pairwise and listwise methods that leverage LLMs to output relative order among documents.
Figure~\ref{fig:different_wise} is a visual aid for each type of method.

\vspace{-0.5em}
\subsection{Pointwise Scoring Approaches}
The scoring approaches employ efficient pointwise inference to individually output relevance scores for each document.
Current scoring methods can be broadly categorized into two types: Query Generation (QG) and Relevance Generation (RG). 

QG \cite{qg, qg-zhuang} treats LLMs as Query Likelihood Models (QLMs) \cite{qlm} and uses the average log-likelihood of generating the query conditioned on a given document as its relevance score. 
\citet{qg-risk} apply Bayesian decision theory to enhance generation under a risk minimization framework, improving the performance of likelihood estimation.
RG directly prompts LLMs to assess the relevance between a single query-document pair and uses the probability of generating specific labels as relevance scores. 
RG-YN \cite{rg-yn} employs a binary classification template \textit{"Does the passage answer the query? Answer \textquotesingle Yes\textquotesingle or \textquotesingle No\textquotesingle "}, and uses the probability of generating \textit{"Yes"} as the relevance score. 
RG-S \cite{rg-s} proposes to adopt fine-grained labels for varying levels of relevance, where numerical labels ranging from $0$ to $k$ have demonstrated superior performance.
MCRanker \cite{on-the-fly} aims to generate relevance scores based on a series of criteria from multiple perspectives, but it heavily depends on pre-defined templates and places significant demands on the capabilities of LLMs.
Moreover, supervised methods like MonoT5 \cite{monot5} and RankT5 \cite{rankt5} present another option, but training such models can be expensive, and acquiring sufficient training data is not always feasible \cite{llm-ir}.

However, previous zero-shot pointwise methods simply evaluate each document in isolation, lacking the comparative insights among candidate documents that could help consistent scoring and accurate ranking.

\vspace{-0.5em}
\subsection{Comparative Approaches}
The comparative approaches prompt LLMs to compare and determine the relative order of input documents, including pairwise and listwise prompting strategies. 

Pairwise methods instruct LLMs to generate the identifier of the document with higher relevance for a pair of documents along with a query. 
To build a global ranking, \citet{pairwise} propose the Allpairs method that considers all pairwise comparisons and uses Copeland's Method \cite{copeland} to calculate final scores. 
Efficient sorting algorithms like heap sort and bubble sort are also employed to speed up the ranking process.
Building on the pairwise prompting approach, \citet{prp-graph} introduce an innovative scoring unit that leverages the generation probability of judgments instead of discrete judgments, and further design a graph-based aggregation approach to obtain a final relevance score for each document. 
Besides, the pairwise comparison can also be utilized as a postprocessing step to adjust the relevance scores generated by the pointwise methods \cite{consolidating}.

Listwise methods instruct LLMs to generate a ranked list of document identifiers according to their relevance to the query.
LRL \cite{lrl} and RankGPT \cite{rankgpt} employ a sliding window strategy to rank a subset of candidate documents each time, alleviating the input length constraints of LLMs.
\citet{found-middle} introduce a permutation self-consistency method, which aggregates multiple generated permutation from multiple shuffled input lists, achieving a more accurate and positionally unbiased ranking.
To mitigate efficiency issues, \citet{tourrank} introduce a tournament-inspired mechanism that ranks documents in a group-based parallel process, and \citet{top-down} propose a parallelizable partitioning algorithm.
Differently, \citet{first} propose to leverage the output logits of the first generated identifier to derive a ranked list with accelerated inference speed.

Comparative methods yield significant performance much better than traditional pointwise methods. 
However, they suffer from severe efficiency, cost, and positional bias issues, making them much less practical.

\begin{figure*}[t]
    \centering
    \includegraphics[width=\textwidth]{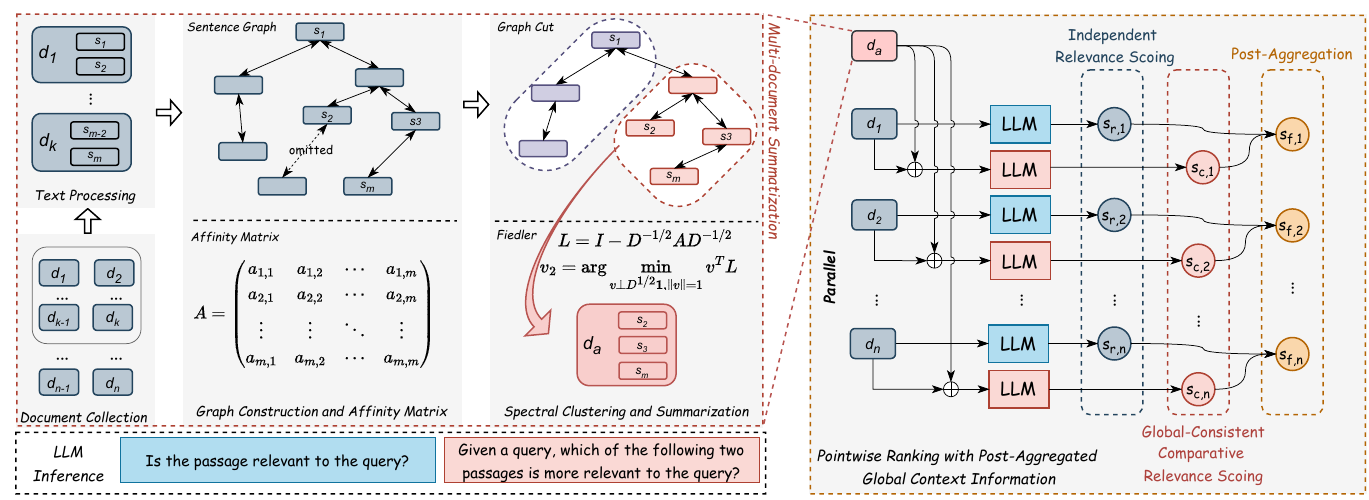}
    \vspace*{-1.2\baselineskip}
    \caption{The overall framework of our proposed pointwise ranking with Post-Aggregated Global Context information (PAGC), including the Global-Consistent Comparative Pointwise ranking strategy (GCCP) and the Post-Aggregation mechanism.}
    \vspace*{-0.8\baselineskip}
    \label{fig:method}
\end{figure*}

\section{Methodology}
\label{sec:method}

\subsection{Preliminaries}
Given a query $q$ and a list of candidate documents $D = (d_{1}, \ldots, d_{n})$, pointwise approaches rely on the relevance score $f(q, d_i) \in \mathbb{R}$ for each query-document pair to rank documents. 
Existing explorations using zero-shot LLMs as pointwise rankers can be broadly divided into two categories: Query Generation (QG)~\cite{qg, qg-zhuang} and Relevance Generation (RG)~\cite{rg-s, rg-yn}.

For QG methods, the relevance score is determined by the average log-likelihood of generating the actual query tokens based on the document:
\begin{equation}
\label{eq:score-qg}
    f_{QG}(q, d_i) = \frac{1}{|q|} \sum_j \text{LLM}(q_j | q_{<j}, d_i, \mathcal{P_{QG}})
\end{equation}
where $|q|$ denotes the token number of query $q$, $d_i$ denotes the document, and $\mathcal{P_{QG}}$ represents the provided prompt for QG.

For RG methods, the relevance score can be calculated based on the log-likelihood of generating specific labels of $k$ different relevance levels via the Expected Relevance value (ER)~\cite{rg-s}:
\begin{equation}
\label{eq:score-rg-s-er}
    f_{RG-SK}(q, d_i) = \sum_{k} \frac{\exp(s_{i,k})}{\sum_{k'} \exp(s_{i,k'})} \cdot k
\end{equation}
where $s_{i,k} = \text{LLM}(l_{k}|q, d_i, \mathcal{P_{RG-K}})$ is the log-likelihood of the LLM generating the label of relevance indicated by $k$.

A widely adopted variant of RG employs a binary relevance judgment, utilizing LLMs to generate tokens "Yes" and "No"~\cite{rg-yn}:
\begin{equation}
\label{eq:score-rg-yn-er}
    f_{RG-YN}(q, d_i) = \frac{\text{exp}(S_{Y})}{\text{exp}(S_{Y}) + \text{exp}(S_{N})}
\end{equation}
where $S_{Y} = s_{i,1} = \text{LLM}(\text{Yes}|q, d_i, \mathcal{P_{RG-YN}})$ and $S_{N} = s_{i,0} = \text{LLM}(\text{No}|q, d_i, \mathcal{P_{RG-YN}})$.

In addition to Expected Relevance value, Peak Relevance likelihood (PR) offers a simpler alternative while maintaining comparable performance\footnote{ER and PR also yielded nearly identical results in our evaluation.}~\cite{rg-s}:
\begin{align}
    \label{eq:max_relevance}
    f_{RG}(q, d_i) = s_{i,k^*} = \text{LLM}(l_{k^*}|q, d_i, \mathcal{P_{RG}})
\end{align}
where $l_{k^*}$ denotes the relevance label with the highest relevance.
PR is more concise and efficient as it relies solely on the probability of the generated label, and can be applied to closed-source LLMs supporting output logits, such as GPT-4. 
Given these advantages, we adopt the PR method to calculate scores for all pointwise methods in our evaluation.

\subsection{Global-Consistent Comparative Pointwise Ranking}

Current pointwise methods evaluate documents independently, leading to potential inconsistencies in relevance ranking due to the absence of inter-document comparisons and global guidance. 
To address these challenges, we propose the Global-Consistent Comparative Pointwise Ranking (GCCP) approach, which integrates pairwise comparisons within the pointwise framework. 
The core idea of GCCP is to establish a reference document, termed the \textit{anchor} document, which reflects the global context of the candidate document set $D$.

\subsubsection{Motivation and Strategy}

In GCCP, the \textit{anchor} document serves as a global reference point, enabling implicit pairwise comparisons within an efficient pointwise ranking framework. 
The key challenge is to construct an \textit{anchor} document with strong discriminative power.
Intuitively, we argue that the \textit{anchor} document should satisfy two essential properties.
\textbf{Query-focused}: the \textit{anchor} should be highly relevant to the query, ensuring that the comparative judgments remain task-oriented and meaningful.
\textbf{Global-representative}: the \textit{anchor} should contain the key information shared among relevant documents, serving as an effective reference point to distinguish documents of varying relevance levels.

Inspired by techniques in Contrastive Learning (CL) for IR systems~\cite{contriever, ance, cllp} and usage of synthetic documents generated by LLMs in IR~\cite{unsup-sigir24, hyde, bias-kdd}, several strategies for building the \textit{anchor} document can be considered:
\begin{itemize}[leftmargin=*]
\item \textbf{Random}: Analogous to the random negative sampling in CL, randomly selecting a document from the candidate set. However, this approach lacks consistency and may not provide a reliable reference point.
\item \textbf{Top}: Similar to the nearest neighbor negative sampling in CL, using the highest-scoring document from initial rankings. While this ensures high query relevance, it may present a limited or biased perspective.
\item \textbf{Synthetic}: Following the trend of using LLMs in IR tasks, generating a synthetic document based on the query. Despite being query-focused, this approach might introduce LLM-specific biases~\cite{bias-kdd} and fail to reflect the actual content distribution in the candidate set.
\item \textbf{Summarization}: Generating a summary that captures key information across candidate documents. This approach potentially balances query relevance with content coverage while avoiding generation biases.
\end{itemize}

We propose to use an unsupervised extractive multi-document summarization (MDS) approach to generate the \textit{anchor} document. This choice is motivated by several factors: (1) it maintains query focus through initial document filtering, (2) it captures distributed information across multiple documents, (3) it avoids potential biases introduced by generative models, and (4) it offers computational efficiency through efficient unsupervised models.

\vspace{-0.5em}
\subsubsection{Spectral-based Multi-Document Summarization}
Inspired by recent successes in spectral methods for multi-document summarization~\cite{spectral-emnlp, spectral-sigir}, we propose a simplified spectral-based MDS algorithm to obtain the \textit{anchor} document. 
The key insight is to leverage spectral graph theory to capture global document structure through local sentence relationships, as illustrated in the upper left part of Figure~\ref{fig:method}.

\vspace{-0.5em}
\paragraph{Sentence Graph Construction}
To effectively model semantic relationships across documents, we construct a sentence-level graph representation.
The process begins by filtering the input to the top-$m$ documents using a preliminary retriever (e.g., BM25), which helps focus on relevant content while maintaining computational efficiency. 
After basic text processing including sentence segmentation and duplicate removal, a similarity graph $G = (V, E)$ is constructed, where vertices $V$ correspond to individual sentences and edges $E$ represent semantic similarities between sentence pairs.

The graph structure is mathematically encoded in an Affinity Matrix $\bm A \in \mathbb{R}^{n \times n}$, where $n$ is the total number of sentences. Each element $a_{i,j}$ is computed as:
\vspace{-0.3em}
\begin{equation}
    a_{i,j} = \begin{cases}
        \cos(e_i, e_j) & \text{if } \cos(e_i, e_j) \geq \theta \\
        0 & \text{otherwise}
    \end{cases}
\vspace{-0.3em}
\end{equation}

where $e_i$ represents the TF-IDF embedding of sentence $s_i$, and $\theta$ is a threshold parameter. 
This thresholding operation produces a sparse affinity matrix that emphasizes strong semantic connections while filtering out weak or noisy relationships, improving the stability of subsequent spectral analysis.

\vspace{-0.5em}
\paragraph{Spectral Analysis}

The spectral analysis begins with the construction of the normalized Laplacian matrix $\bm L$:
\begin{equation}
\bm L = \bm I - \bm D^{-1/2}\bm A\bm D^{-1/2}
\end{equation}
where $\bm I$ is the identity matrix and $\bm D$ is the degree matrix with diagonal elements $d_{ii} = \sum_j a_{ij}$. The normalized Laplacian $\bm L$ captures both local connectivity patterns and global graph structure.

The central component of the spectral analysis is the Fiedler vector $\bm v_2$, which is the eigenvector corresponding to the second smallest eigenvalue $\lambda_2$ of $\bm L$. This vector can be obtained by solving:
\vspace{-0.3em}
\begin{equation}
\bm L\bm v_2 = \lambda_2 \bm v_2
\vspace{-0.3em}
\end{equation}

The Fiedler vector provides an optimal solution to the normalized cut problem~\cite{fiedler}, which aims to partition the graph into two cohesive subgroups while minimizing the inter-group connections. Mathematically, this is equivalent to solving:
\vspace{-0.3em}
\begin{equation}
\bm v_2 = \arg\min_{\bm v \perp \bm D^{1/2}\mathbf{1}, \|\bm v\|=1} v^T \bm L v
\end{equation}
\vspace{-0.3em}

This optimization effectively separates sentences into two clusters: those central to the overall content and those that are peripheral. The sign of each component in $\bm v_2$ indicates to which cluster a sentence belongs, providing a natural partitioning mechanism.

\vspace{-0.5em}
\paragraph{Anchor Document Generation}
The final step involves constructing the \textit{anchor} document using the spectral analysis results. 
Based on the Fiedler vector, we partition the sentences into two clusters according to the sign of their corresponding components. 
The cluster containing more sentences is selected as the source for the \textit{anchor} document, as it represents the more common themes across the document set.

To preserve discourse coherence, sentences within the selected cluster are reordered according to their original positions in their respective source documents. 
The \textit{anchor} document is then constructed by selecting a predetermined number of sentences from this ordered sequence. Formally, let $C = \{s_1, s_2, ..., s_{|C|}\}$ be the set of sentences in the larger cluster, ordered by their original document positions. The \textit{anchor} document $d_a$ is defined as:
\vspace{-0.3em}
\begin{equation}
d_a = \{s_i \in C : i \leq z\}
\vspace{-0.3em}
\end{equation}

where $z$ is the predefined number of sentences to include in the \textit{anchor}. 
This approach provides an efficient way to generate an \textit{anchor} document that maintains both content representation and structural coherence. 
By focusing on the larger cluster and preserving the original sentence order, the generated \textit{anchor} document effectively captures the core information while maintaining the natural flow of the source documents, serving as a reliable reference point for our GCCP framework.

\vspace{-0.5em}
\subsubsection{Contrastive Relevance Scoring}
With the \textit{anchor} document $d_a$ generated, the contrastive relevance score $f(q, d_i, d_a)$ can be computed as:
\vspace{-0.3em}
\begin{equation}
\label{eq:score-gccp}
    f_c(q, d_i, d_a) = \text{LLM}(d_i|q, d_i, d_a, \mathcal{P_{GCCP}})
\vspace{-0.3em}
\end{equation}
where $\mathcal{P_{GCCP}}$ is a prompt designed to elicit a comparative judgment between $d_i$ and $d_a$ in the context of query $q$. 
We adopt the prompt used in pairwise methods~\cite{pairwise, prp-graph} as $\mathcal{P_{GCCP}}$, that is \textit{"Given a query, which of the following two passages is more relevant to the query?"}.

\vspace{-0.5em}
\subsection{Post-Aggregation}
The contrastive relevance scores are then effectively integrated with traditional pointwise scores $f(q, d_i)$ derived from either QG or RG methods (as outlined in Equations \ref{eq:score-qg}, \ref{eq:score-rg-s-er}, and \ref{eq:score-rg-yn-er}). The final combined score can be obtained through a simple linear aggregation scheme:
\vspace{-0.3em}
\begin{equation}
\label{eq:score-pagc}
f_{\text{final}}(q, d_i) = \frac{1}{|\mathcal{R}|+1}(\sum_{\mathcal{R}} f(q, d_i) + f_c(q, d_i, d_a))
\vspace{-0.3em}
\end{equation}
where $\mathcal{R}$ is the set of independent pointwise scoring methods.

This refinement addresses the initial challenges by incorporating collection-wide context into individual document scores, providing a basis for implicit pairwise comparisons against a common reference, and potentially reducing inconsistencies in relevance judgments across the document set. 
The GCCP approach thus combines the efficiency of pointwise ranking with the contextual awareness of pairwise comparisons, potentially leading to more consistent and accurate document rankings in zero-shot scenarios.

\section{Experiments}

\subsection{Experimental Setup}

\subsubsection{Datasets and Evaluation Metrics}

We evaluate our method on two popular benchmark datasets: TREC DL \cite{dl19, dl20} and BEIR \cite{beir}.
\textbf{TREC DL} is a prominent benchmark widely used in IR research. 
We utilize the test sets from TREC DL 19 and TREC DL 20, comprising 43 and 54 queries, respectively.
\textbf{BEIR} is a heterogeneous zero-shot evaluation benchmark that includes a variety of retrieval tasks across multiple domains. 
Consistent with prior studies \cite{rankgpt, setwise, pairwise, prp-graph}, our evaluation leverages 8 datasets: \textit{Covid}, \textit{Touche}, \textit{DBPedia}, \textit{SciFact}, \textit{Signal}, \textit{News}, \textit{Robust04}, and \textit{NFCorpus}.

The effectiveness is evaluated using NDCG@10 \cite{ndcg}. 
To evaluate efficiency, the latency per query is recorded, and the API cost per query is estimated based on GPT-4-o calculations.

\vspace{-0.5em}
\subsubsection{Comparison Methods}
We conduct a comparison of our methods against several state-of-the-art baselines in zero-shot document ranking, categorized into \textbf{Listwise}, \textbf{Pairwise}, and \textbf{Pointwise} approaches. 
Among the Listwise methods, \textbf{RankGPT} \cite{rankgpt} generates a global order for input documents using a sliding window strategy, while \textbf{Setwise} \cite{setwise} leverages sorting algorithms like heapsort to focus on top-k ranking by selecting the most relevant document from candidate sets. For Pairwise approaches, the \textbf{PRP} \cite{pairwise} method produces relative relevance labels through document pair comparisons in Allpair and heapsort versions, and \textbf{PRP-Graph} \cite{prp-graph} quantifies comparison certainty using output probabilities with a parameter setting of $r=40$. 
In the category of Pointwise methods, \textbf{QG} \cite{qg} assesses documents based on query generation likelihood, and \textbf{RG-YN} \cite{rg-yn} employs "Yes/No" likelihood for scoring. Further, \textbf{RG-S(0,4)} \cite{rg-s} evaluates relevance on a numeric scale with $k=4$. Our contributions include \textbf{GCCP}, which integrates global-consistent pairwise comparisons, and \textbf{PAGC}, a post-aggregation technique that blends various pointwise strategies. Notably, the combination termed as \textsf{PAGC-QSG} encompasses \textsf{\underline{Q}G}, \textsf{RG-\underline{S}(0,4)}, and \textsf{\underline{G}CCP} methods.

\vspace{-0.5em}
\subsection{Implementation Details}
Pyserini~\cite{pyserini} is utilized with default settings to generate the initial top-100 BM25 first-stage ranking for all datasets. 
Consistent with prior studies \cite{setwise, prp-graph}, we employ open-source Flan-t5 LLMs \cite{flan-models}, including Flan-t5-large (780M), Flan-t5-xl (3B). 
We also report the results of Flan-ul2 \cite{ul2}, which has been extensively evaluated in other studies \cite{pairwise, consolidating, rg-s}. 
We set $m=10$ and $z=10$ to construct the \textit{anchor} document. 
All relevance scores for pointwise methods are calculated using the PR likelihood described in Eq~\ref{eq:max_relevance}. 
For PAGC, linear aggregation is implemented. 
Prompts used in this paper follow \citet{prompt-zhuang}. 
Experiments are conducted on a local GPU workstation equipped with four NVIDIA RTX A6000 48GB GPUs.

\begin{table*}[t]
\centering
\caption{The performance of different pointwise models and previous pointwise models post-aggregated with our GCCP on the TREC DL and BEIR benchmarks. All the models re-rank BM25 top 100 documents and the effectiveness is evaluated by nDCG@10. The best result is in bold and the second-best is marked with an underline for each backbone LLM. "$\dagger$" indicates statistically significantly better than the corresponding single model with Paired t-test $p \leq 0.05$.}
\vspace*{-0.5\baselineskip}
\renewcommand\arraystretch{0.7}
\begin{threeparttable}
\resizebox{\linewidth}{!}{
      \begin{tabular}{c|l|cc|c|cccccccc|c}
      \toprule
      &  & \multicolumn{3}{c|}{\textbf{TREC DL}} & \multicolumn{9}{c}{\textbf{BEIR}} \\
      \midrule
      & \multicolumn{1}{c|}{\textbf{Methods}} & \textbf{dl19} & \textbf{dl20} & \textbf{Avg} & \textbf{covid} & \textbf{robust04} & \textbf{touche} & \textbf{scifact} & \textbf{signal} & \textbf{news} & \textbf{dbpedia} & \textbf{nfcorpus} & \textbf{Avg} \\
      \midrule
            & BM25  & 0.5058 & 0.4796 & 0.4927 & 0.5947 & 0.4070 & 0.4422 & 0.6789 & 0.3304 & 0.3952 & 0.3180 & 0.3218 & 0.4360 \\
      \midrule
      \multirow{7}[8]{*}{ \rotatebox[origin=c]{90}{Flan-t5-large}}    
            & GCCP & 0.6480 & 0.5671 & 0.6076 & \underline{0.7693} & 0.4427 & 0.2730 & 0.5871 & 0.2955 & \underline{0.4338} & \underline{0.4251} & 0.3504 & 0.4471 \\
      \cmidrule{2-14}  
            & QG        & 0.5548 & 0.5653 & 0.5601 & 0.6636 & 0.4384 & 0.2597 & \underline{0.6451} & \textbf{0.3163} & 0.4128 & 0.3035 & 0.3366 & 0.4220 \\
            & \phantom{00}+ GCCP   & 0.6787\tnote{$\dagger$} & 0.6152\tnote{$\dagger$} & 0.6470\tnote{$\dagger$} & \textbf{0.7906}\tnote{$\dagger$} & 0.4720\tnote{$\dagger$} & \textbf{0.2933}\tnote{$\dagger$} & \textbf{0.6691}\tnote{$\dagger$} & \underline{0.3127} & \textbf{0.4494}\tnote{$\dagger$} & \textbf{0.4298}\tnote{$\dagger$} & \underline{0.3616}\tnote{$\dagger$} & \textbf{0.4723}\tnote{$\dagger$} \\
      \cmidrule{2-14}     
            & RG-YN & 0.6643 & 0.6150 & 0.6397 & 0.6884 & 0.4605 & 0.2479 & 0.5635 & 0.2823 & 0.3691 & 0.3478 & 0.3349 & 0.4118 \\
            & \phantom{00}+ GCCP & \textbf{0.7012}\tnote{$\dagger$} & \textbf{0.6281}\tnote{$\dagger$} & \textbf{0.6647}\tnote{$\dagger$} & 0.7641\tnote{$\dagger$} & \underline{0.4914}\tnote{$\dagger$} & \underline{0.2928}\tnote{$\dagger$} & 0.6145\tnote{$\dagger$} & 0.3027\tnote{$\dagger$} & 0.4112\tnote{$\dagger$} & 0.4181\tnote{$\dagger$} & \textbf{0.3638}\tnote{$\dagger$} & 0.4573\tnote{$\dagger$} \\
      \cmidrule{2-14}        
            & RG-S(0,4) & 0.6291 & 0.5806 & 0.6049 & 0.7246 & 0.4789 & 0.2163 & 0.5690 & 0.2952 & 0.3588 & 0.3299 & 0.3162 & 0.4111 \\
            & \phantom{00}+ GCCP & \underline{0.6887}\tnote{$\dagger$} & \underline{0.6273}\tnote{$\dagger$} & \underline{0.6580}\tnote{$\dagger$} & 0.7633\tnote{$\dagger$} & \textbf{0.5008}\tnote{$\dagger$} & 0.2712\tnote{$\dagger$} & 0.6319\tnote{$\dagger$} & 0.3109\tnote{$\dagger$} & 0.4319\tnote{$\dagger$} & 0.4200\tnote{$\dagger$} & 0.3555\tnote{$\dagger$} & \underline{0.4607}\tnote{$\dagger$} \\
      \midrule
      \multirow{7}[8]{*}{ \rotatebox[origin=c]{90}{Flan-t5-xl}}  
            & GCCP & \underline{0.7020} & 0.6672 & 0.6846 & 0.7689 & 0.5262 & \underline{0.3052} & 0.6812 & 0.3175 & \textbf{0.4746} & 0.4261 & 0.3715 & 0.4839 \\
      \cmidrule{2-14} 
            & QG    & 0.5397 & 0.5418 & 0.5408 & 0.6787 & 0.4262 & 0.2196 & \underline{0.6944} & 0.3015 & 0.4211 & 0.3088 & 0.3454 & 0.4245 \\
            & \phantom{00}+ GCCP & 0.6933\tnote{$\dagger$} & 0.6767\tnote{$\dagger$} & 0.6850\tnote{$\dagger$} & \underline{0.7757}\tnote{$\dagger$} & 0.5251\tnote{$\dagger$} & \textbf{0.3056}\tnote{$\dagger$} & \textbf{0.7002} & 0.3251\tnote{$\dagger$} & 0.4674\tnote{$\dagger$} & \textbf{0.4290}\tnote{$\dagger$} & \textbf{0.3797}\tnote{$\dagger$} & \textbf{0.4885}\tnote{$\dagger$} \\
      \cmidrule{2-14}
            & RG-YN & 0.6730 & 0.6493 & 0.6612 & 0.7173 & 0.5048 & 0.2838 & 0.6427 & 0.3109 & 0.4502 & 0.3198 & 0.3546 & 0.4480 \\
            & \phantom{00}+ GCCP & 0.6969\tnote{$\dagger$} & \underline{0.6810}\tnote{$\dagger$} & \underline{0.6890}\tnote{$\dagger$} & \textbf{0.7761}\tnote{$\dagger$} & \underline{0.5346}\tnote{$\dagger$} & 0.3016\tnote{$\dagger$} & 0.6854\tnote{$\dagger$} & 0.3230\tnote{$\dagger$} & 0.4740\tnote{$\dagger$} & 0.4149\tnote{$\dagger$} & 0.3756\tnote{$\dagger$} & 0.4857\tnote{$\dagger$} \\
      \cmidrule{2-14}
            & RG-S(0,4) & 0.6844 & 0.6501 & 0.6673 & 0.7154 & 0.5143 & 0.2758 & 0.6662 & 0.3166 & 0.4627 & 0.3587 & 0.3546 & 0.4580 \\
            & \phantom{00}+ GCCP & \textbf{0.7022}\tnote{$\dagger$} & \textbf{0.6880}\tnote{$\dagger$} & \textbf{0.6951}\tnote{$\dagger$} & 0.7685\tnote{$\dagger$} & \textbf{0.5353}\tnote{$\dagger$} & 0.3043\tnote{$\dagger$} & 0.6892\tnote{$\dagger$} & \textbf{0.3254}\tnote{$\dagger$} & 0.4731\tnote{$\dagger$} & 0.4256\tnote{$\dagger$} & 0.3725\tnote{$\dagger$} & 0.4867\tnote{$\dagger$} \\
      \midrule
      \multirow{7}[8]{*}{ \rotatebox[origin=c]{90}{Flan-ul2}}
            & GCCP   & 0.7155 & \textbf{0.7040} & 0.7098 & 0.7729 & 0.5320 & 0.2920 & 0.6983 & 0.3075 & 0.4961 & \textbf{0.4396} & \underline{0.3792} & 0.4897 \\
      \cmidrule{2-14}
            & QG    & 0.5662 & 0.5686 & 0.5674 & 0.7029 & 0.4511 & 0.2341 & 0.6989 & 0.2945 & 0.4103 & 0.3291 & 0.3449 & 0.4332 \\
            & \phantom{00}+ GCCP & 0.7146\tnote{$\dagger$} & 0.6986\tnote{$\dagger$} & 0.7066\tnote{$\dagger$} & \underline{0.7794}\tnote{$\dagger$} & 0.5340\tnote{$\dagger$} & \underline{0.3018}\tnote{$\dagger$} & \textbf{0.7357}\tnote{$\dagger$} & 0.3189\tnote{$\dagger$} & \textbf{0.5078}\tnote{$\dagger$} & 0.4327\tnote{$\dagger$} & \textbf{0.3830}\tnote{$\dagger$} & \underline{0.4992}\tnote{$\dagger$} \\
      \cmidrule{2-14}
            & RG-YN & 0.6866 & 0.6678 & 0.6772 & 0.7300 & 0.5394 & 0.2718 & 0.7098 & \underline{0.3280} & 0.4598 & 0.3808 & 0.3656 & 0.4732 \\
            & \phantom{00}+ GCCP & \underline{0.7206}\tnote{$\dagger$} & 0.7023\tnote{$\dagger$} & \underline{0.7115}\tnote{$\dagger$} & 0.7781\tnote{$\dagger$} & \underline{0.5467} & 0.2930\tnote{$\dagger$} & 0.7295\tnote{$\dagger$} & 0.3186 & 0.4864\tnote{$\dagger$} & 0.4265\tnote{$\dagger$} & \underline{0.3792}\tnote{$\dagger$} & 0.4948\tnote{$\dagger$} \\
      \cmidrule{2-14}
            & RG-S(0,4) & 0.6830 & 0.6701 & 0.6766 & 0.7585 & 0.5396 & 0.2776 & 0.7298 & \textbf{0.3298} & 0.4778 & 0.4168 & 0.3722 & 0.4878 \\
            & \phantom{00}+ GCCP & \textbf{0.7227}\tnote{$\dagger$} & \underline{0.7030}\tnote{$\dagger$} & \textbf{0.7129}\tnote{$\dagger$} & \textbf{0.7905}\tnote{$\dagger$} & \textbf{0.5479}\tnote{$\dagger$} & \textbf{0.3157}\tnote{$\dagger$} & \underline{0.7325} & 0.3208 & \underline{0.5010}\tnote{$\dagger$} & \underline{0.4370}\tnote{$\dagger$} & 0.3779 & \textbf{0.5029}\tnote{$\dagger$} \\
      \bottomrule
      \end{tabular}%
}
\vspace*{-0.5\baselineskip}
\end{threeparttable}
\label{tab:main-pt}%
\end{table*}%
  
\subsection{Experimental Results}
\subsubsection{Compare with Scoring Approaches}

Table~\ref{tab:main-pt} illustrates the performance of different pointwise models on the TREC DL and BEIR benchmarks. 
We compare our proposed GCCP model against prior pointwise methods, including QG, RG-YN, and RG-S(0,4), as well as these methods when post-aggregated with GCCP. 
From the results, several observations can be made:

Firstly, when deployed independently, our proposed GCCP consistently surpasses previous pointwise methods on both benchmarks across all three models, with the sole exception occurring on TREC DL with the Flan-t5-large model. 
Notably, on the BEIR benchmark, GCCP implemented with Flan-t5-large achieves an average nDCG@10 score of 0.4471, significantly outperforming previously recorded scores of 0.4220, 0.4118, and 0.4111. 
Similarly, with the Flan-t5-xl model, GCCP obtains a score of 0.4839, markedly better than scores of 0.4245, 0.4480, and 0.4580 achieved by prior models.

Secondly, the integration of GCCP with all three existing pointwise models results in notable improvements, highlighting the effectiveness of our method for post-aggregating global context information. 
For instance, when GCCP is combined with the QG model under the Flan-t5-large configuration, the average nDCG@10 score on TREC DL rises from 0.5601 to 0.6470. 
A similar trend of improvement can be observed across different configurations, consistently enhancing ranking performance.
Our methods establish new state-of-the-art performance levels among pointwise ranking approaches on two benchmarks.

\vspace{-0.5em}
\subsubsection{Compare with Comparative Approaches}

\begin{table*}[htbp]
\centering
\caption{The performance of our post-aggregated pointwise models against comparative models on the TREC DL and BEIR benchmarks. All the models re-rank BM25 top 100 documents and the effectiveness is evaluated by nDCG@10. The best result is in bold and the second-best is marked with an underline for each backbone LLM.}
\vspace*{-0.5\baselineskip}
\renewcommand\arraystretch{0.8}
\begin{threeparttable}
\resizebox{\linewidth}{!}{
      \begin{tabular}{c|c|cc|c|cccccccc|c}
      \toprule
      &   & \multicolumn{3}{c|}{\textbf{TREC DL}} & \multicolumn{9}{c}{\textbf{BEIR}} \\
      \midrule      
      & \textbf{Methods} & \textbf{dl19} & \textbf{dl20} & \textbf{Avg} & \textbf{covid} & \textbf{robust04} & \textbf{touche} & \textbf{scifact} & \textbf{signal} & \textbf{news} & \textbf{dbpedia} & \textbf{nfcorpus} & \textbf{Avg} \\
      \midrule
            & BM25  & 0.5058 & 0.4796 & 0.4927 & 0.5947 & 0.4070 & 0.4422 & 0.6789 & 0.3304 & 0.3952 & 0.3180 & 0.3218 & 0.4360 \\
      \midrule
      \multirow{7}[4]{*}{ \rotatebox[origin=c]{90}{Flan-t5-large}} 
            & RankGPT & 0.6690 & 0.6260 & 0.6475 & 0.7560 & 0.4750 & \textbf{0.3270} & 0.6390 & 0.3080 & \underline{0.4530} & \underline{0.4440} & 0.3340 & 0.4670 \\
            & PRP-Allpair   & 0.6646 & 0.6201 & 0.6424 & \underline{0.7716} & 0.4462 & 0.2887 & 0.6515 & 0.3188 & 0.4219 & 0.4299 & 0.3417 & 0.4588 \\
            & PRP-Heapsort   & 0.6570 & 0.6190 & 0.6380 & 0.7610 & 0.4020 & \underline{0.3180} & \underline{0.6710} & \underline{0.3250} & 0.4400 & 0.4140 & 0.3360 & 0.4584 \\
            & PRP-Graph-40 & 0.6690 & 0.6270 & 0.6480 & \textbf{0.7788} & 0.4837 & 0.3024 & 0.6355 & \textbf{0.3391} & \textbf{0.4593} & \textbf{0.4474} & 0.3422 & \underline{0.4736} \\
            & Setwise-Heapsort & 0.6700 & 0.6180 & 0.6440 & 0.7680 & 0.4620 & 0.3030 & 0.6200 & 0.3190 & 0.4390 & 0.4130 & 0.3250 & 0.4561 \\
      \cmidrule{2-14}           
            & PAGC-QYG & \textbf{0.7005}  & \textbf{0.6521}  & \textbf{0.6763}  & 0.7657  & \underline{0.4988}  & 0.2944  & 0.6568  & 0.3150  & 0.4324  & 0.4220  & \textbf{0.3675}  & 0.4691  \\
            & PAGC-QSG & \underline{0.6973}  & \underline{0.6433}  & \underline{0.6703}  & 0.7700  & \textbf{0.5105}  & 0.2753  & \textbf{0.6728}  & 0.3218  & 0.4518  & 0.4291  & \underline{0.3603}  & \textbf{0.4740}  \\
      \midrule
      \multirow{7}[4]{*}{ \rotatebox[origin=c]{90}{Flan-t5-xl}}   
            & RankGPT & 0.6890 & 0.6720 & 0.6805 & 0.7360 & 0.5260 & 0.3100 & 0.6860 & 0.3200 & 0.4720 & \textbf{0.4490} & 0.3600 & 0.4824 \\
            & PRP-Allpair   & \textbf{0.7108} & 0.6821 & 0.6965 & \underline{0.7802} & \underline{0.5467} & 0.2955 & 0.7002 & 0.3220 & 0.4819 & 0.4295 & 0.3665 & 0.4903 \\
            & PRP-Heapsort   & 0.7050 & \textbf{0.6920} & \textbf{0.6985} & 0.7780 & \textbf{0.5500} & 0.3030 & \underline{0.7110} & 0.3170 & 0.4710 & 0.4170 & 0.3550 & 0.4878 \\
            & PRP-Graph-40 & 0.6972 & 0.6767 & 0.6870 & 0.7756 & 0.5381 & \textbf{0.3122} & \textbf{0.7249} & 0.3079 & \underline{0.4853} & \underline{0.4339} & 0.3546 & \underline{0.4916} \\
            & Setwise-Heapsort & 0.6930 & 0.6780 & 0.6855 & 0.7570 & 0.5200 & 0.2830 & 0.6770 & 0.3140 & 0.4650 & 0.4280 & 0.3520 & 0.4745 \\
      \cmidrule{2-14}            
            & PAGC-QYG & 0.6932  & \underline{0.6884}  & 0.6908  & \textbf{0.7806}  & 0.5339  & 0.3099  & 0.7011  & \textbf{0.3345}  & \textbf{0.4870}  & 0.4158  & \textbf{0.3810}  & \textbf{0.4930}  \\
            & PAGC-QSG & \underline{0.7068}  & 0.6872  & \underline{0.6970}  & 0.7722  & 0.5355  & \underline{0.3112}  & 0.7041  & \underline{0.3270}  & 0.4742  & 0.4310  & \underline{0.3776}  & 0.4916  \\
      \midrule
      \multirow{7}[3]{*}{ \rotatebox[origin=c]{90}{Flan-ul2}}   
            & RankGPT  & 0.7132 & 0.6993 & 0.7063 & 0.7902 & 0.5395 & \textbf{0.3291} & 0.7334 & 0.3217 & \textbf{0.5152} & 0.4306 & 0.3709 & 0.5038 \\
            & PRP-Allpair   & 0.7198 & \underline{0.7045} & \textbf{0.7122} & \textbf{0.8042} & \underline{0.5510} & 0.3011 & \underline{0.7463} & 0.3233 & 0.4795 & \textbf{0.4577} & 0.3790 & \underline{0.5053} \\
            & PRP-Heapsort   & 0.7143 & 0.7015 & 0.7079 & 0.7709 & 0.5296 & 0.3007 & 0.7402 & \textbf{0.3330} & 0.5078 & \underline{0.4402} & 0.3776 & 0.5000 \\
            & PRP-Graph-40 & 0.7110 & \textbf{0.7082} & 0.7096 & \underline{0.7945} & 0.5309 & 0.3006 & 0.7253 & 0.3159 & 0.4943 & 0.4318 & 0.3742 & 0.4959 \\
            & Setwise-Heapsort & \underline{0.7205} & 0.7001 & 0.7103 & 0.7715 & 0.5283 & \underline{0.3169} & 0.7307 & \underline{0.3277} & \underline{0.5127} & 0.4324 & 0.3755 & 0.4995 \\
      \cmidrule{2-14}           
            & PAGC-QYG & \textbf{0.7215}  & 0.6980  & 0.7098  & 0.7827  & \textbf{0.5511}  & 0.3001  & 0.7382  & 0.3228  & 0.5006  & 0.4257  & \textbf{0.3815}  & 0.5003  \\
            & PAGC-QSG & 0.7200  & 0.7029  & \underline{0.7115}  & 0.7904  & 0.5486  & 0.3096  & \textbf{0.7474}  & 0.3235 & 0.5097 & 0.4382 & \underline{0.3810} & \textbf{0.5061} \\
      \bottomrule
      \end{tabular}
}
\vspace*{-0.5\baselineskip}
\end{threeparttable}
\label{tab:main-comparative}%
\end{table*}%

Table~\ref{tab:main-comparative} presents the performance comparison between various zero-shot ranking models on the TREC DL and BEIR benchmarks. 
We evaluate our proposed PAGC model against existing comparative methods, encompassing both pairwise and listwise approaches. 
Notably, \textsf{PAGC-QSG} denotes the aggregation of \textsf{\underline{Q}G}, \textsf{RG-\underline{S}(0,4)}, and \textsf{\underline{G}CCP} methods, which would be \textsf{QG+RG-S(0,4)+GCCP} in Table~\ref{tab:main-pt}.
The experimental results reveal several noteworthy findings:

Firstly, unlike some pointwise methods shown in Table~\ref{tab:main-pt} that underperform BM25, our proposed PAGC, and other comparative methods consistently achieve significantly superior average performance across both benchmarks compared to the BM25 baseline.

Secondly, while maintaining computational efficiency, our method demonstrates competitive performance against more complex and computationally intensive comparative approaches. 
Notably, when conducted under the Flan-t5-large configuration, our approach achieves substantial improvements on the TREC DL benchmark, surpassing the previous state-of-the-art PRP-Graph-40 score of 0.6480 with an improved score of 0.6763.
Additionally, our method consistently attains the highest average performance on the BEIR benchmark across model sizes.

\vspace{-0.5em}
\subsubsection{Theoretical Analysis of Efficiency}

\begin{table}[h]
\centering
\caption{The theoretical lowest time complexity of various methods and the number of LLM forward passes needed in the worst case and per query on TREC DL. ($N$: number of documents to rank. $r$: number of repeats. $s$: step size for sliding window. $k$: number of top-$k$ documents to find. $c$: number of compared documents at each step.)}
\vspace*{-0.5\baselineskip}
\renewcommand\arraystretch{0.8}
\resizebox{0.9\linewidth}{!}{%
    \begin{tabular}{cccc}
    \toprule
    \textbf{Methods} & \textbf{Time Complexity} & \textbf{\#LLM calls} & \textbf{\#Inference}\\ 
    \midrule
        QG, RG & {$O(1)$} & {$O(N)$} & 100 \\ 
    \midrule
        RankGPT & $O(N)$ & $O(r * (N/s))$ & 245 \\ 
    \midrule
        PRP-Allpair & {$O(1)$} & $O(N^2 -N)$ & 9900 \\ 
    \midrule
        PRP-Heapsort & $O(k * log_{2}N)$ & $O(k * log_{2}N)$ & 243.1 \\ 
    \midrule
        Setwise-Heapsort & $O(k * log_{c}N)$ & $O(k * log_{c}N)$ & 128.7 \\ 
    \midrule
        GCCP (Ours) & {$O(1)$} & {$O(N)$} & 100 \\ 
    \midrule
        PAGC (Ours) & {$O(1)$} & $O(2N)$, $O(3N)$ & 200, 300  \\ 
    \bottomrule
    \end{tabular}%
}
\vspace*{-0.8\baselineskip}
\label{tab:effi-analysis}
\end{table}


Table~\ref{tab:effi-analysis} provides a comparative analysis of various methods with respect to their theoretical minimum time complexity, the number of LLM forward passes required in the worst-case scenario, and the actual number of LLM inferences performed per query on TREC DL.

Our proposed methods, GCCP and PAGC, demonstrate optimal time complexity at $O(1)$, aligning with pointwise approaches such as QG and RG. This efficiency surpasses all comparative methods, which operate at time complexities of $O(N)$ or $O(\log N)$. In terms of LLM calls, GCCP maintains an $O(N)$ complexity, while PAGC requires $O(2N)$ or $O(3N)$, reflecting its ability to adapt to different performance thresholds.
Although PRP-Heapsort and Setwise-Heapsort theoretically offer beneficial complexities at $O(\log N)$, their actual inferences suggest limitations in practical performance. This arises from the small input sizes typical in LLM applications, where $N$ is relatively small and $k$ is not insignificantly smaller. The lack of parallelism further constrains these methods. While Setwise-Heapsort performs slightly fewer inferences at 128.7 compared to PAGC, it shows considerably lower effectiveness as highlighted in Table~\ref{tab:main-pt}. Moreover, PRP-Allpair requires $N^2 - N$ inferences, which significantly impacts its practical feasibility.

Our methods achieve a commendable balance between theoretical efficiency and real-world performance, providing a more robust solution for complex ranking tasks.

\vspace{-0.5em}
\subsubsection{Effectiveness and Efficiency Trade-offs}

\begin{figure*}[ht]
    \begin{subfigure}{1\columnwidth}
        \centering
        \includegraphics[width=\linewidth]{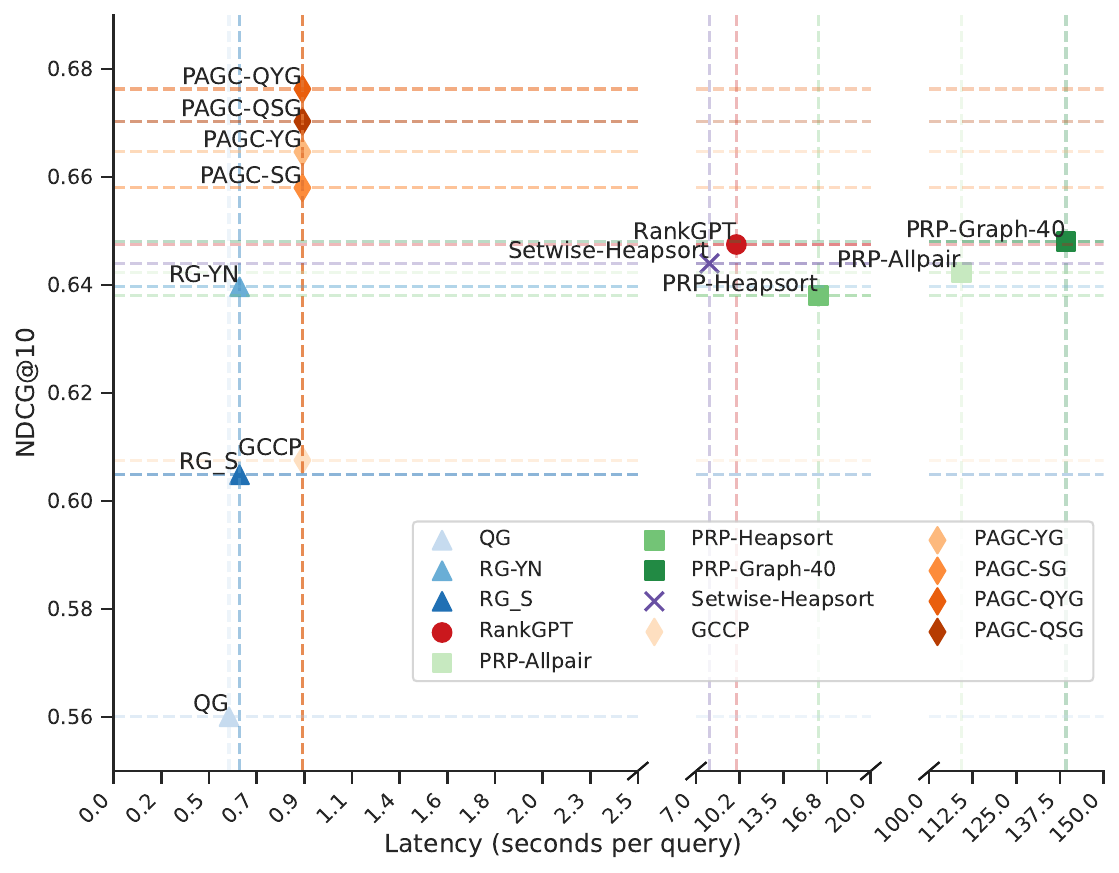}
        \caption{Latency v.s. nDCG@10}
        \label{fig:latency}
    \end{subfigure}
    \begin{subfigure}{1\columnwidth}
        \centering
        \includegraphics[width=\linewidth]{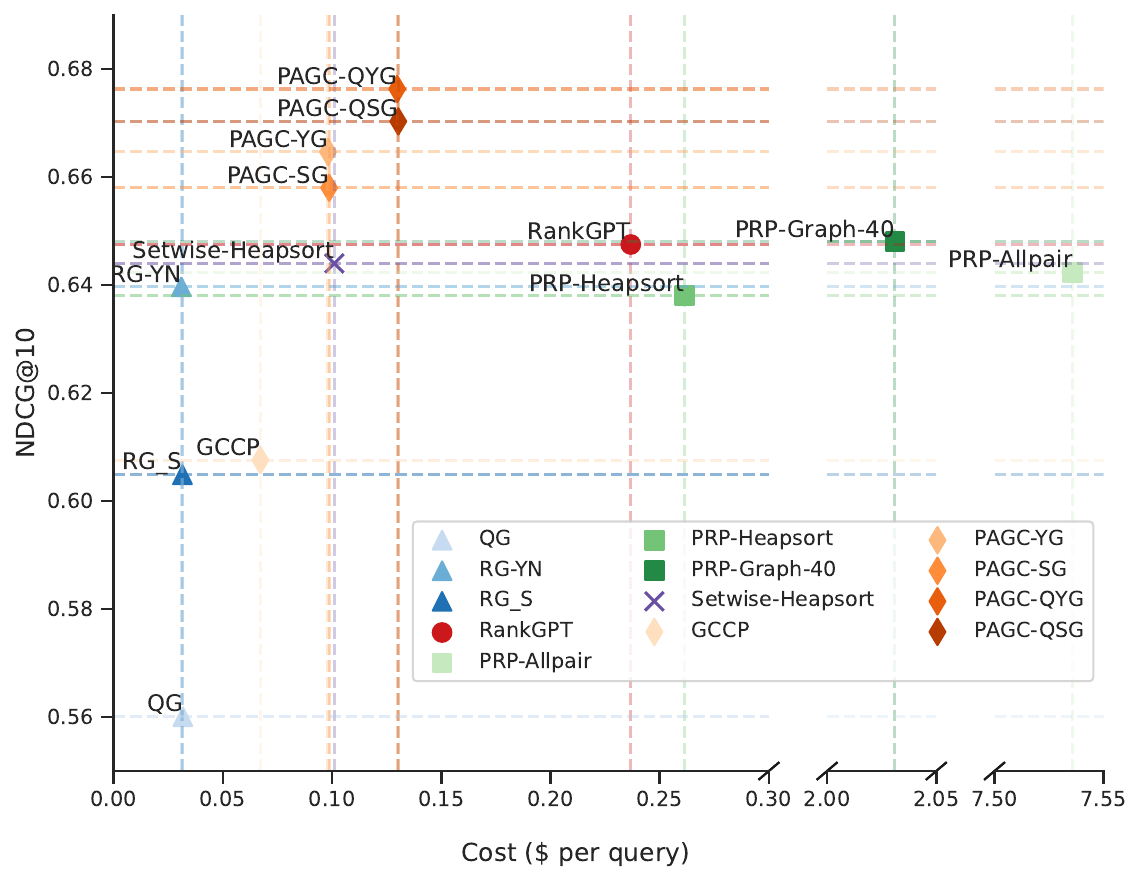}
        \caption{Cost v.s. nDCG@10}
        \label{fig:cost}
    \end{subfigure}
    \caption{Relationship between Cost / Latency and NDCG@10 on the TREC DL benchmark. (a) shows the trade-off between latency and nDCG@10, while (b) illustrates the relationship between cost and nDCG@10. These plots demonstrate how different approaches balance effectiveness and efficiency.}
    \label{fig:tradeoffs}
\end{figure*}

We compare the estimated cost and latency per query of our proposed GCCP and PAGC against several baseline methods on the TREC DL benchmark using Flan-t5-large. Our efficiency evaluation settings are aligned with those used in \cite{setwise}. The cost estimation is based on OpenAI's GPT-4-o pricing structure\footnote{At the time of writing, OpenAI's costs were \$0.0025 per 1,000 prompt tokens and \$0.01 per 1,000 generated tokens.}. For the PAGC method, which aggregates multiple pointwise rankings, latency is calculated as the maximum of its components since all components can execute simultaneously.

Figure~\ref{fig:tradeoffs} provides an intuitive illustration of the relationship between cost, latency, and performance across various methods. Previous pointwise methods are associated with the lowest actual cost and latency. However, their NDCG@10 scores are noticeably below those of other methods, except for RG-YN, which achieves comparable results. Comparative methods, on the other hand, demonstrate significantly higher latency and cost. For instance, PRP-Allpair incurs a cost of approximately \$7.5 per query, whereas pointwise methods require no more than \$0.15. Even the most efficient among the comparative methods, RankGPT, almost doubles the cost compared to pointwise solutions. In terms of latency, PRP-Graph-40 is the least efficient, requiring about 137 seconds per query, while pointwise methods take less than 1 second.

Ultimately, our methods demonstrate that, with only a slight increase in latency of about 0.3 seconds and an additional cost of approximately \$0.6, they achieve significantly higher NDCG@10 scores than previous pointwise methods and even outperform comparative methods.

\vspace{-0.5em}
\subsection{More Analyses on PAGC}
\subsubsection{Aggregation Method}

We investigate the impact of different rank aggregation methods on the ranking performance of PAGC. 
Utilizing various techniques available in PyFlagr\footnote{\url{https://flagr.site/}}, we incorporate methods such as Linear, Borda~\cite{broda}, Condorcet, Copeland, Outrank~\cite{outrank}, MCT~\cite{mct}, and DIBRA~\cite{dibra} into our analysis. 
For this purpose, we use the models PAGC-QYG and PAGC-QSG, reporting their average NDCG@10 scores across the TREC DL and BEIR benchmarks for each aggregation method with Flan-t5-large.

\begin{figure}[h]
    \centering
    \includegraphics[width=\columnwidth]{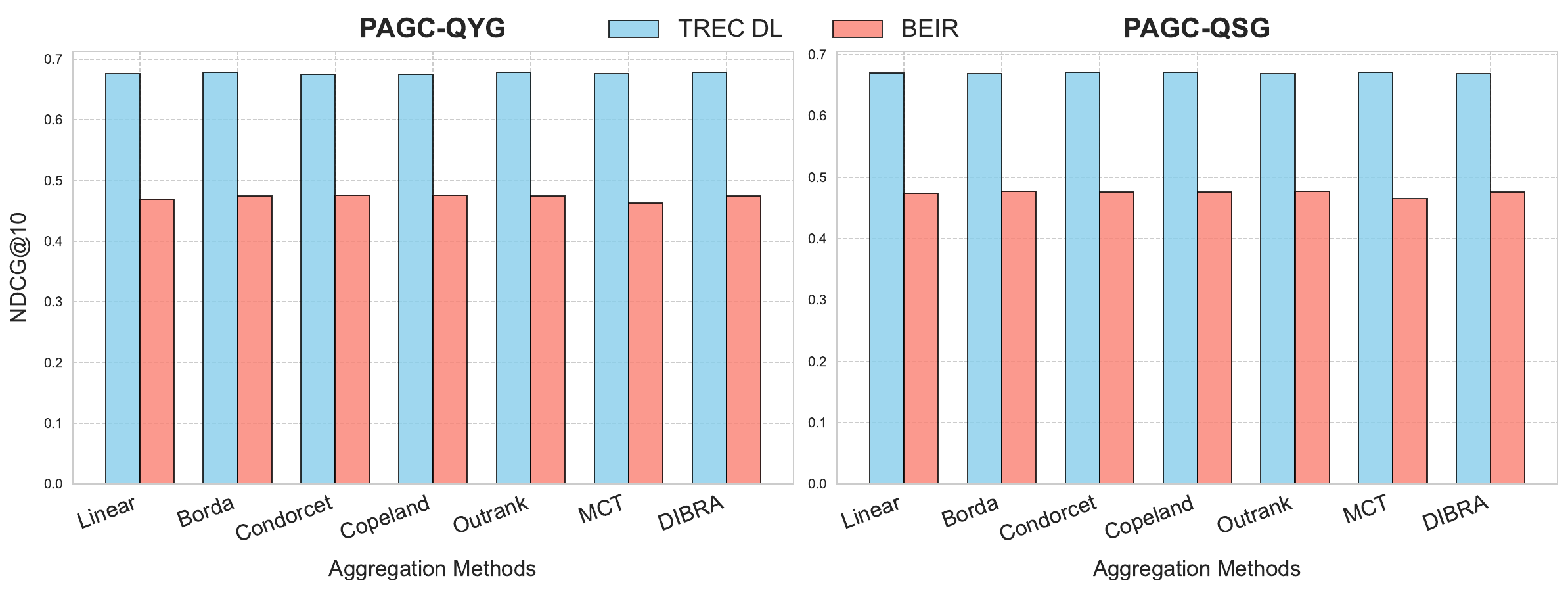}
    \vspace*{-2\baselineskip}
    \caption{Performance results of PAGC with different rank aggregation methods.}
    \vspace*{-0.5\baselineskip}
    \label{fig:different_agg}
    \vspace*{-0.5\baselineskip}
\end{figure}

As depicted in Figure~\ref{fig:different_agg}, all aggregation approaches effectively enhance ranking performance, with nearly identical NDCG@10 scores observed regardless of whether score-based methods like Linear or positional methods such as Borda and Outrank are used. 
This consistency highlights the robustness of pointwise methods across different aggregation strategies. 
Consequently, our method adopts the highly efficient linear aggregation approach, achieving substantial performance improvements with minimal aggregation cost.

\vspace{-0.5em}
\subsubsection{Impact of $m$}
\label{sec:different_m}

We then experiment to examine the impact of $m$, the number of candidates used to construct the \textit{anchor}. 
Specifically, we use the top-$m$ candidates from the initial ranking as the input set for MDS, varying $m$ from 10 to 100 in increments of 10. 
Our experiments include four individual pointwise methods and their various aggregations, averaging the NDCG@10 scores on both the TREC DL and BEIR benchmarks with Flan-t5-large.

\begin{figure}[h]
    \centering
    \vspace*{-0.8\baselineskip}
    \includegraphics[width=\columnwidth]{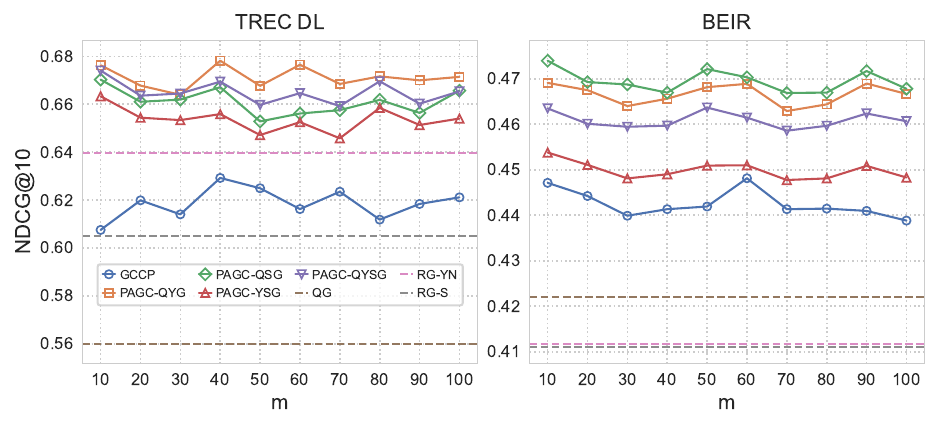}
    \vspace*{-2\baselineskip}
    \caption{Performance results of GCCP and PAGC with different values of $m$.}
    \vspace*{-0.5\baselineskip}
    \label{fig:different_top_m}
    \vspace*{-0.3\baselineskip}
\end{figure}

As shown in Figure~\ref{fig:different_top_m}, PAGC-QYG achieves the best performance across nearly all $m$ values on the TREC DL benchmark, whereas PAGC-QSG excels on the BEIR benchmark. 
For PAGC, almost optimal performance is observed at $m=10$ on both datasets, with performance generally declining as $m$ increased, except for a few instances where performance is comparable to $m=10$. 
This pattern is also evident for GCCP on the BEIR dataset, though it does not hold on TREC DL, possibly due to the smaller number of test queries in TREC DL, which could lead to greater random variability.

\subsubsection{Heterogeneous Aggregation}

\begin{table}[t]
    \centering
    \caption{Comparison of homogeneous and heterogeneous aggregation components.}
    \renewcommand\arraystretch{0.9}
    \resizebox{\linewidth}{!}{%
        \begin{tabular}{clcccccc}
        \toprule
        & \multirow{2}{*}{\textbf{Method}} & \multicolumn{2}{c}{\textbf{Large}} & \multicolumn{2}{c}{\textbf{XL}} & \multicolumn{2}{c}{\textbf{UL2}} \\
        \cmidrule(lr){3-4} \cmidrule(lr){5-6} \cmidrule(lr){7-8}
        & & TREC & BEIR & TREC & BEIR & TREC & BEIR \\
        \midrule
        \multirow{6}[2]{*}{ \rotatebox[origin=c]{90}{Two Comp.}} & \multicolumn{7}{c}{\textit{Homogeneous}} \\
        & RG-S+RG-YN & 0.6301 & 0.4159 & 0.6582 & 0.4575 & 0.6820 & 0.4845 \\
        & RG-S+QG & \underline{0.6406} & 0.4460 & 0.6565 & 0.4669 & 0.6811 & 0.4855 \\
        & GCCP $\times$ 2 & 0.6350 & \underline{0.4551} & \underline{0.6950} & \textbf{0.4874} & \underline{0.7103} & \underline{0.4942} \\
        & \multicolumn{7}{c}{\textit{Heterogeneous}} \\
        & RG-S+GCCP & \textbf{0.6580} & \textbf{0.4607} & \textbf{0.6951} & \underline{0.4867} & \textbf{0.7129} & \textbf{0.5029} \\
        \midrule
        \multirow{7}[2]{*}{ \rotatebox[origin=c]{90}{Three Comp.}} & \multicolumn{7}{c}{\textit{Homogeneous}} \\
        & RG-S+RG-YN+QG & 0.6457 & 0.4352 & 0.6654 & 0.4699 & 0.6847 & 0.4902 \\
        & GCCP $\times$ 3 & 0.6446 & 0.4609 & \underline{0.6954} & \underline{0.4917} & \underline{0.7099} & 0.4987 \\
        & \multicolumn{7}{c}{\textit{Heterogeneous}} \\
        & RG-S+RG-YN+GCCP & 0.6634 & 0.4538 & 0.6926 & 0.4860 & 0.7084 & \underline{0.5019} \\
        & RG-S+QG+GCCP & \underline{0.6703} & \textbf{0.4740} & \textbf{0.6970} & 0.4916 & \textbf{0.7115} & \textbf{0.5061} \\
        & RG-YN+QG+GCCP & \textbf{0.6763} & \underline{0.4691} & 0.6908 & \textbf{0.4930} & 0.7098 & 0.5003 \\
        \bottomrule
        \end{tabular}%
    }
\label{tab:agg_sources}
\end{table}

To further verify the effectiveness of incorporating global context information through post-aggregation, we conduct experiments on different aggregation components. Specifically, we examine the performance of homogeneous pointwise method aggregations versus heterogeneous aggregations across three model configurations. We evaluate their average NDCG@10 scores on the TREC DL and BEIR benchmarks, where heterogeneous approaches are differentiated by previous independent scoring and our comparative scoring methods.

As presented in Table~\ref{tab:agg_sources}, the results indicate that heterogeneous aggregation generally achieves superior outcomes across different scenarios. While the GCCP method alone provides the best single ranking performance (Table~\ref{tab:main-pt}), using multiple instances of GCCP (as discussed in \S\ref{sec:different_m}, e.g., GCCP with different values of $m$ such as GCCP $\times$ 2 or GCCP $\times$ 3) does not surpass the results obtained from combining GCCP with other weaker but heterogeneous pointwise methods. This highlights the significance of leveraging global context information through complementary heterogeneity, which leads to improved performance compared to homogeneous combinations.

\vspace{-0.5em}
\subsubsection{Impact of Anchor Strategy}

Table~\ref{tab:anchor_type} presents an evaluation of various strategies for constructing the \textit{anchor} document within the GCCP and PAGC models, showcasing NDCG@10 and the number of LLM calls required per query on the TREC DL and BEIR benchmarks. 
The assessed strategies include our Spectral-based MDS approach, a Top-ranked candidate document, a Randomly selected document, and a Synthetic document generated using GPT-4-o with prompts from Hyde\cite{hyde}. 
The "Top" strategy involves randomly selecting five documents from the top-10 candidates, each serving as an anchor, with the performance averaged for these selections. 
The "Random" strategy follows a similar process but selects from the entire document set. Variants with the suffix "-5" involve aggregating the results from five anchors with those from other pointwise methods.

\begin{table}[t]
\centering
\caption{Performance results of different \textit{anchor} construction methods.}
\vspace*{-0.5\baselineskip}
\renewcommand\arraystretch{0.9}
\resizebox{0.9\linewidth}{!}{%
    \begin{tabular}{cccccc}
    \toprule
    \textbf{Method} & \textbf{Variant} & \textbf{TREC DL} & \textbf{BEIR} & \textbf{\#LLM calls} \\ 
    \midrule
    \multirow{3}{*}{GCCP} 
        & Spectral & \underline{0.6076} & \textbf{0.4471} & 100 \\ 
        & Top & \textbf{0.6099} & \underline{0.4346} & 100 \\ 
        & Random & 0.6061 & 0.4253 & 100 \\ 
        & Synthetic & 0.5785 & 0.4121 & 101 \\ 
    \midrule
    \multirow{5}{*}{PAGC-QYG} 
        & Spectral & \textbf{0.6763} & \textbf{0.4691} & 200 \\ 
        & Top & \underline{0.6702} & 0.4630 & 200 \\ 
        & Random & 0.6660 & 0.4575 & 200 \\ 
        & Synthetic & 0.6491 & 0.4417 & 201 \\ 
        & Top-5 & 0.6685 & \underline{0.4689} & 600 \\ 
        & Random-5 & 0.6664 & 0.4672 & 600 \\ 
    \midrule
    \multirow{5}{*}{PAGC-QSG} 
        & Spectral & \textbf{0.6703} & \textbf{0.4740} & 200 \\ 
        & Top & \underline{0.6693} & 0.4645 & 200 \\ 
        & Random & 0.6678 & 0.4616 & 200 \\ 
        & Synthetic & 0.6502 & 0.4483 & 201 \\ 
        & Top-5 & 0.6666 & \underline{0.4718} & 600 \\ 
        & Random-5 & 0.6673 & 0.4666 & 600 \\ 
    \bottomrule
    \end{tabular}%
}
\label{tab:anchor_type}
\end{table}

The Spectral method consistently exhibits strong performance across both datasets. Specifically, within the PAGC-QYG and PAGC-QSG models, the Spectral variant achieves the highest NDCG@10 scores on both TREC DL (0.6763 and 0.6703, respectively) and BEIR (0.4691 and 0.4740, respectively), demonstrating its effectiveness in utilizing global context information, thus optimizing anchor selection.
Conversely, the Top method, while also performing well, generally falls slightly short of the Spectral method, except in the GCCP model on the TREC DL benchmark, where it marginally outperforms the Spectral method with a score of 0.6099. The Random method yields lower scores, with a maximum of 0.6660 on TREC DL in the PAGC-QYG model, indicating the inadequacy of random selection as a strategic approach.
The synthetic approach shows inferior results. The GCCP's contrastive relevance score with a synthetic anchor is less variably distributed, often exhibiting high concentration within narrow ranges, leading to poorer outcomes. This could be attributed to biases introduced by generative LLMs, as documented in \cite{bias-kdd}, and potentially due to discrepancies in word distribution between LLM-generated documents and original documents.
Variants such as Top-5 and Random-5, which involve using multiple anchors, do not significantly improve performance despite requiring more LLM calls (600). This suggests that a more targeted anchor selection with fewer anchors, as seen in the Spectral method, is both efficient and effective.

These findings emphasize the importance of strategic \textit{anchor} construction with our proposed Spectral-based MDS method, affirming its role in enhancing ranking performance by effectively incorporating global context information.

\section{Conclusion}
\label{sec:conclusion}

This paper addresses the challenge of inconsistent scoring in traditional pointwise ranking methods by introducing a Global-Consistent Comparative Pointwise Ranking (GCCP) strategy. 
By employing an anchor document to effectively incorporate global context, GCCP enhances scoring consistency in zero-shot scenarios. 
Additionally, the Post-Aggregation with Global Context (PAGC) framework allows for seamless integration of enhanced scores with existing methodologies, further boosting effectiveness without sacrificing efficiency. 
Extensive evaluation on TREC DL and BEIR benchmarks demonstrates that our approach consistently outperforms traditional pointwise methods, highlighting its potential as a scalable and robust solution for modern ranking challenges involving LLMs.


\newpage
\bibliographystyle{ACM-Reference-Format}
\balance
\bibliography{sample-base}



\end{document}